\documentclass[%
 reprint,
 amsmath,amssymb,
 aps,aip
]{revtex4-1}

\pdfoutput=1

\usepackage{graphicx}
\usepackage{dcolumn}
\usepackage{bm}
\usepackage{subfigure}

\usepackage{color}
\usepackage{algorithm}
\usepackage{algorithmic}

\usepackage{CJK}

\begin{document}


\title{ Machine learning study of the relationship between the geometric and entropy discord}
\author{Qin-Sheng Zhu $^{1}$}
\email{zhuqinsheng@gmail.com}
\author{ Xiao-Yu Li $^{2}$}%
\author{Ming-Zheng Zhu $^{2}$}%
\author{Yi-Ming Huang $^{2}$}%
\author{Hao Wu $^{1}$}%
\author{Shao-Yi Wu $^{1}$}%
\affiliation{%
  $^{1}$School of Physics, University of Electronic Science and Technology of China, Chengdu, 610054, P.R.China
}%

\affiliation{%
  $^{2}$School of information and software engineering, University of Electronic Science and Technology of China, Chengdu, 610054, P.R.China
}%

\date{\today}

\begin{abstract}
As an important resource to realize quantum information, quantum correlation displays different behaviors, freezing phenomenon and non-localization,
which are dissimilar to the entanglement and classical correlation, respectively. In our setup, the ordering of the value of quantum correlation is represented for different quantization
methods by considering an open quantum system scenario. The machine learning method (neural network method) is then adopted to train for the construction of a bridge between the R\`{e}nyi discord ($\alpha=2$) and the geometric discord (Bures distance) for $X$ form states. Our results clearly demonstrate that the machine learning method is useful for studying the differences and commonalities of different quantizing methods of quantum correlation.
\begin{description}
\item[PACS numbers]
03.65.Yz, 03.67.-a, 89.70.-a, 03.65.Ud.
\item[Key words]
machine learning; quantum correlation.
\end{description}
\end{abstract}

\pacs{Valid PACS appear here}
\maketitle


\section{\label{sec:level1}Introduction}

With the development of the quantum technology, many novel instruments and ideas arise to serve for people's life, such as quantum communication and quantum computer [1].
When we deal with the composite quantum systems in these applications, the superposition principle which is a basic theory of quantum mechanics and the tensorial structure of the Hilbert space have been widely applied to describe these composite quantum systems. The concept of entanglement [2] which is a kind of special superposition states is always naturally involved. The earliest researches that entanglement is equivalent to quantum correlation had been regarded as reasonable for many years. Simultaneously, many different quantification methods have been put forward in this period, including geometric [3] and entropy methods (the most famous is concurrence [4] for the entanglement of two partial system ), and many interesting properties of entanglement had been found for different quantum systems, such as sudden death and sudden spring [5-7].

About twenty years ago, Ollivier and Zurek [8] and Henderson and Vedral [9] introduced the concept of "quantum discord" . It told us that the entanglement does not account for all nonclassical correlations and that even the states with zero entanglement  usually contain quantum correlations[8,10]. And there is an universal consensus that entanglement entirely captures quantum correlation only for a global pure state [11]. So, many related works have been presented for X states [12-13] and some open quantum systems [14-23] in the past few years, and the unique freezing phenomenon is found, which reveals a robust feature of a family of two-qubit models subject to nondissipative decoherence [11,24-26].

In the experimental implementation perspective, the R\`{e}nyi entropy
\begin{eqnarray}
S_{\alpha}(\rho)=\frac{1}{1-\alpha}\log Tr[\rho ^{\alpha}]
\end{eqnarray}
which arouses much attention in recent years because the R\`{e}nyi entropy shows quantitative bounds for different parameter $\alpha$ comparing the von Neumann entropy, and  easier implement than the von Neumann entropy for measuring entanglement [27-28].  Here the parameter $\alpha\in(0,1)\cup(1,\infty)$ and the logarithm
is in base 2. Notably, the R$\acute{e}$nyi entropy will reduce to the von Neumann entropy when $\alpha\rightarrow 1$.
As an natural extension of quantum discord, the R\`{e}nyi entropy discord (RED)[29-30] is also put forward.
Therefore, it is valuable to study the properties and the role of RED in quantum information field.

From the geometric viewpoint, several other quantization methods are proposed, such as Hilber-Schmidt[31] ($D_{HS}$), Bures distance [3,11,32-33]($D_{Br}$), trace-norm and Hellinger [11,34]($D_{HL}$). Unlike the quantum discord, the geometric discords quantify the quantum correlation by searching the minimum distance between the quantum states and zero quantum correlation states, and showed the classification of quantum states, such as classical states, quantum-classical states and quantum states. Based on these classification, it is better to understand the difference between the entanglement and the quantum correlation by geometric definition [3,11]. Simultaneously, the different quantization methods are ordered as:
\begin{eqnarray}
&&D_{Br}(\rho)\geq \{D_{HL}(\rho), D_{HS}(\rho)\}\nonumber\\
&&D_{Br}(\rho)\geq E(\rho)
\end{eqnarray}
which is discussed in Ref.[11,33]. Here, $E(\rho)$ denotes the geometric quantification of the entanglement. Unfortunately, the concurrence can be smaller or larger than quantum discord.

Although a series of works mentioned the above properties of geometric and entropy discords, respectively, no clearly relationship between the geometric and entropy style discords has been established because all the discords are defined by the complex nonlinear mathematical forms. The new idea to solve this problem is finding a method to obtain the relation only based on partial data (the value of quantum correlation of partial quantum states). As part of both artificial intelligence and statistics,  machine learning come from the computer science field in which the goal is to learn the potential patterns from prior given data sets, and make a decision or prediction for future unknown situation based on this learned patterns. Recently, these learning tools have been used for dealing with some quantum problems, such as quantum state tomography [35], and quantum many-body problem [36]. These results suggest that machine learning can be a new platform for solving some problems of quantum physics. In addition, establishing the relationship between geometric and entropy discords through machine learning method will be beneficial to reveal some hidden physical character of the quantum state, for example, which quantum states can exist the freezing phenomenon under the same condition for different discords [37].

In this work, we calculate the value of entanglement and different discords for two qubit open system under the $X$ form initial states, and show the order of these value. Notably, our result not only gives the powerful proof for Ref.[11,33], but also firstly answers whether the RED of $\alpha=2$ can resolve the problem-\emph{``quantum discord can be larger or smaller than the concurrence"}.
Furthermore, another highlight of this work is constructing the relationship between $D_{Br}$ and the RED of $\alpha=2$ by the use of machine learning method.

\emph{\textbf{The ordering of the value of quantum  correlation for different quantization methods.}}

Stemming from the research works of Ollivier and Zurek [8] and Henderson and Vedral [9], many efforts have been devoted to study the quantum correlation for different systems by use of different methods. Ref.[3] shows that the geometric measures have a nice ordering feature (seeing Eq.(2)). In contrast, the concurrence (entanglement) can be larger or smaller than the quantum discord [22, 38]. Now, whether the R\`{e}nyi discord is larger than the concurrence is still unknown, although it is monotone increasing with $\alpha$ [29-30](for $\alpha=1$, the R\`{e}nyi discord  reduce to the quantum discord).

In what follows, we consider an open quantum system scenario, and study the ordering of quantum correlation under different quantification methods.

Here, we consider the anisotropic coupling two qubit system which is coupled to two correlated Fermi-spin environments, respectively. The Hamiltonian of the total system has the following form:
\begin{eqnarray}
H&=&H_{s}+\sum_{i=1,2}(H_{E_{i}}+H_{sE_{i}})+q S_{1}^{z}S_{2}^{z}\nonumber\\
H_{s}&=&J_{1}(\sigma_{1}^{x}\sigma_{2}^{x}+\sigma_{1}^{y}\sigma_{2}^{y})+J_{2}\sigma_{1}^{z}\sigma_{2}^{z}+\sum_{i=1,2}\omega_{i}\sigma_{i}^{z}\nonumber\\
H_{E_{i}}&=&\alpha_{i}S_{i}^{z};H_{sE_{i}}=\gamma_{i}\sigma_{i}^{z}S_{i}^{z}
\end{eqnarray}

where, $J_{1}$ and $J_{2}$ are the anisotropic coupling parameters between two spin particles. $\omega_{i}$ and $\alpha_{i}$ are the frequencies of spin particle and environmental spin particle, respectively. $q $ describes an Ising-type correlation between the environments. $S_{i}^{z}=\sum_{k=1}^{N_{i}}\frac{\sigma_{z}^{k,i}}{2}$ is the collective spin operators, $\sigma_{z}^{k,i}$ are the Pauli matrices and each environment $E_{i}$ consists of $N_{i}$ particles with spin $1/2$.

Here, the states $|j,m\rangle$ denote the orthogonal bases in the environment Hilbert space $H_{B}$ which satisfy [22,39]:
\begin{eqnarray}
S^{2}|j,m\rangle&=&j(j+1)|j,m\rangle;\nonumber\\
S^{z}|j,m\rangle&=&m|j,m\rangle;S^{2}=(S^{x})^{2}+(S^{y})^{2}+(S^{z})^{2}\nonumber\\
j&=&0,...,\frac{N}{2}; m=j,...,-j\nonumber\\
\nonumber
\end{eqnarray}
For the initial state $\rho(0)=\rho_{s}(0)\otimes\rho_{E}(0)$ condition, the reduced density matrices $\rho_{s}(t)$ of the system can be obtained
\begin{eqnarray}
\rho_{d}(t)&=&\frac{1}{Z}\sum_{j_{1}=0}^{N_{1}/2}\sum_{m_{1}=-j_{1}}^{j_{1}}\sum_{j_{2}=0}^{N_{2}/2}\sum_{m_{2}=-j_{2}}^{j_{2}}
\frac{\nu(N_{1},j_{1})\nu(N_{2},j_{2})}{e^{\beta q m_{1}m_{2}}e^{\beta\alpha_{1}m_{1}}e^{\beta\alpha_{2}m_{2}}}\nonumber\\
&&\times V^{\dagger}U^{\dagger}(t)\rho^{'}_{s}(0)U(t) V
\end{eqnarray}
where $V^{\dagger}=[\langle00|\langle10|\langle01|\langle11|]$
\begin{eqnarray}
&&U(t)=\nonumber\\
&&\left[
\begin{array}{cccc}
 e^{-iE_{1}t} &0 &0 &0\nonumber\\
 0 &\frac{e^{-iE_{2}t}Q_{1}-e^{-iE_{3}t}Q_{2}}{Q_{1}-Q_{2}} &\frac{2J_{1}(e^{-iE_{2}t}-e^{-iE_{3}t})}{Q_{1}-Q_{2}}&0\nonumber\\
 0 &\frac{2J_{1}(e^{-iE_{2}t}-e^{-iE_{3}t})}{Q_{1}-Q_{2}} &\frac{-e^{-iE_{2}t}Q_{2}+e^{-iE_{3}t}Q_{1}}{Q_{1}-Q_{2}} &0\nonumber\\
 0 &0 &0 &e^{-iE_{4}t} \nonumber\\
\end{array}
\right]\nonumber\\
\nonumber
\end{eqnarray}

where $|0\rangle$ and $|1\rangle$ denote the spin up and down states, respectively. So the two spin particles consist $\mathbb{C}^{2} \bigotimes \mathbb{C}^{2}$ Hilbert space
and the state space can be expanded by the basis $[|00\rangle,|01\rangle,|10\rangle,|11\rangle]$. $Q_{1,2}=Q\pm\sqrt{Q^{2}+4J_{1}^{2}}$ $Q=\omega_{1}+\gamma_{1}m_{1}-\omega_{2}-\gamma_{2}m_{2}$,$E_{1}=\omega_{1}+\gamma_{1}m_{1}+\omega_{2}+\gamma_{2}m_{2}+J_{2}$,$E_{2,3}=-J_{2}\pm\sqrt{Q^{2}+4J_{1}^{2}}$ and $E_{4}=-E_{1}+2J_{2}$. $\rho_{s}^{'}(0)$ is the matrix form of $\rho_{s}^{'}(0)$ under the basis $V$.

Notice, when the initial density matrix $\rho_{s}^{'}(0)$ has X form [12-13]
\begin{eqnarray}
\rho_{s}'(0)=\left[
\begin{array}{cccc}
 a     & 0 &0 &\delta \\
 0    & b &\beta &0\\
 0    & \beta^{*}  &c &0\\
 \delta^{*}      & 0  &0  &d\\
\end{array}
\right]
\end{eqnarray}
which satisfies $a,b,c,d\geq0$,$ a+b+c+d=1$, $||\delta||^{2}\leq ad$ and $||\beta||^{2}\leq bc$, it is easy to check that the matrix form of $\rho_{s}(t)$ is also X form.

In Fig.1, it is shown the difference between the different quantification methods of quantum correlation for $X$ initial states (total 65880 states) generated by our considered open quantum system, including $D_{HS}$, $D_{HL}$, $D_{Br}$, $RED$ and $concurrence$. The value of each dot denotes the value of difference between the different quantification methods at different time.  Noticing, this ordering is only between geometric methods ($D_{HS}$, $D_{HL}$, $D_{Br}$) or entropy style methods ($RED$, $concurrence$). The parameters are $\alpha_{1}=250 ps^{-1}$, $\alpha_{2}=200 ps^{-1}$, $\omega_{1}=5ps^{-1},\omega_{2}=6ps^{-1}$,$q=30ps^{-1}$, $J_{1}=9ps^{-1}$, $J_{2}=11ps^{-1}$ ,$\beta=1/77$, $N_{1}=14$, $N_{2}=12$, $\gamma_{1}=0.2ps^{-1}$ and $\gamma_{2}=0.3ps^{-1}$. Fig1. (a)-(b) further prove the reliability of the relationship $D_{Br}(\rho)\geq \{D_{HL}(\rho), D_{HS}(\rho)\}$. In contrast, Fig1.(c) shows that $D_{HL}$ can be larger or smaller than the $D_{HS}$. Simultaneously, for entropy style quantization of quantum correlation, RED is still larger or smaller than the $concurrence$ for $\alpha=2$ condition, as shown in Fig.1(d). Combined with the monotone increasing of RED with $\alpha$, we obtained that the $RED$ shows a better value ordering of quantum correlation than quantum discord (RED reduce to quantum discord when $\alpha\rightarrow 1$) for $\alpha > 1$. As a conclusion, the value ordering of quantum correlation of different quantization methods in Fig.1 shows that $D_{Br}$ [3,11,32-33] and $RED$ [29-30] quantization methods can be regarded as better than the others methods.

\begin{figure}
\begin{minipage}[t]{0.5\linewidth}
\centering
\includegraphics[width=1.6in]{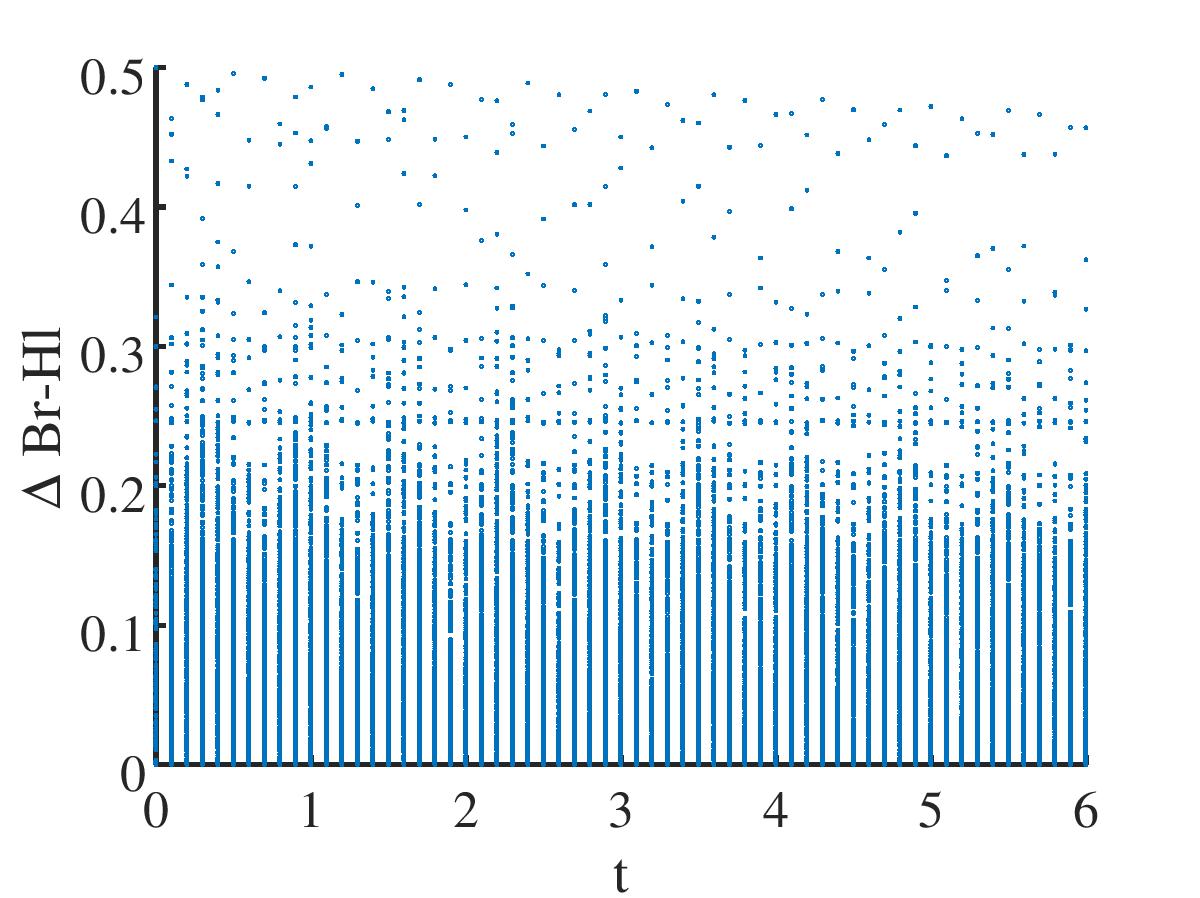}
\label{fig:side:a}

\end{minipage}%
\begin{minipage}[t]{0.5\linewidth}
\centering
\includegraphics[width=1.6in]{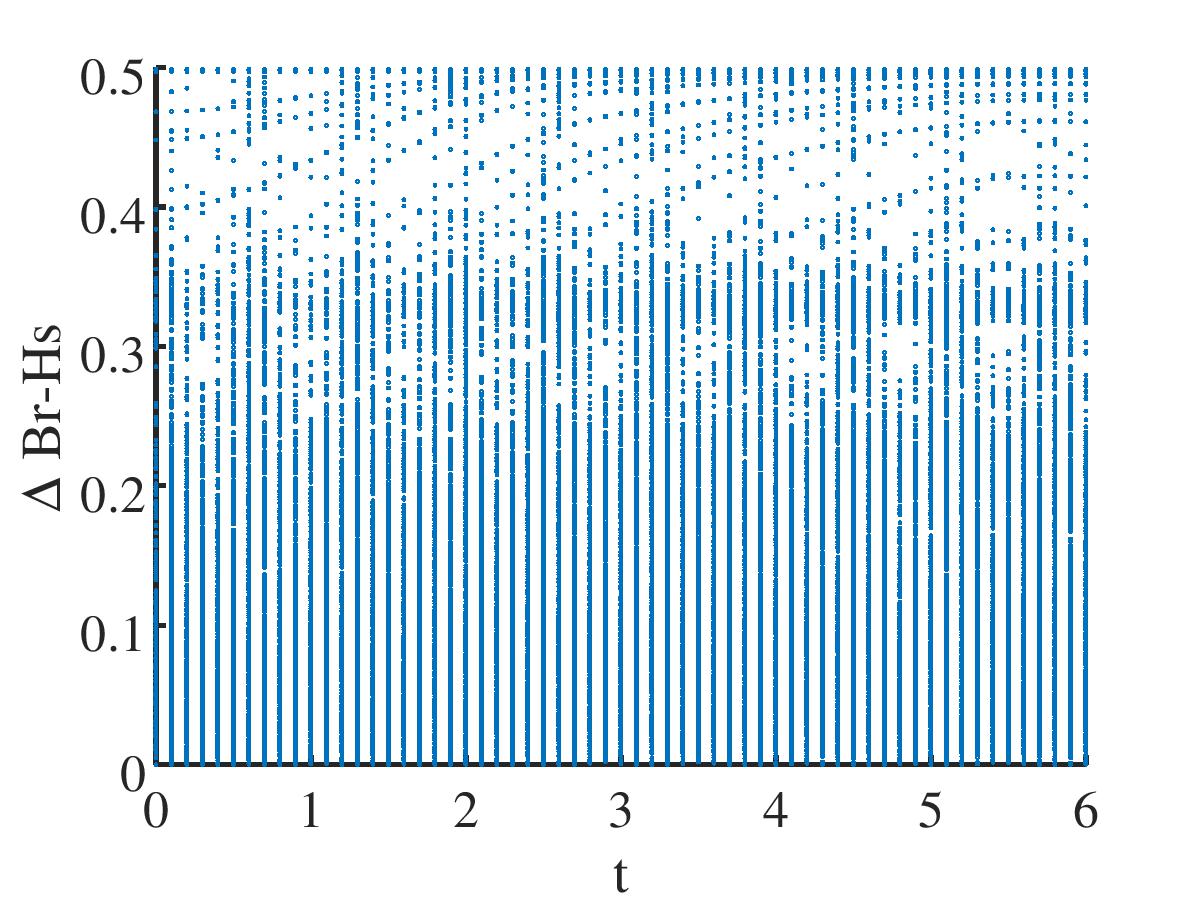}
\label{fig:side:b}

\end{minipage}
\begin{minipage}[t]{0.5\linewidth}
\centering
\includegraphics[width=1.56in]{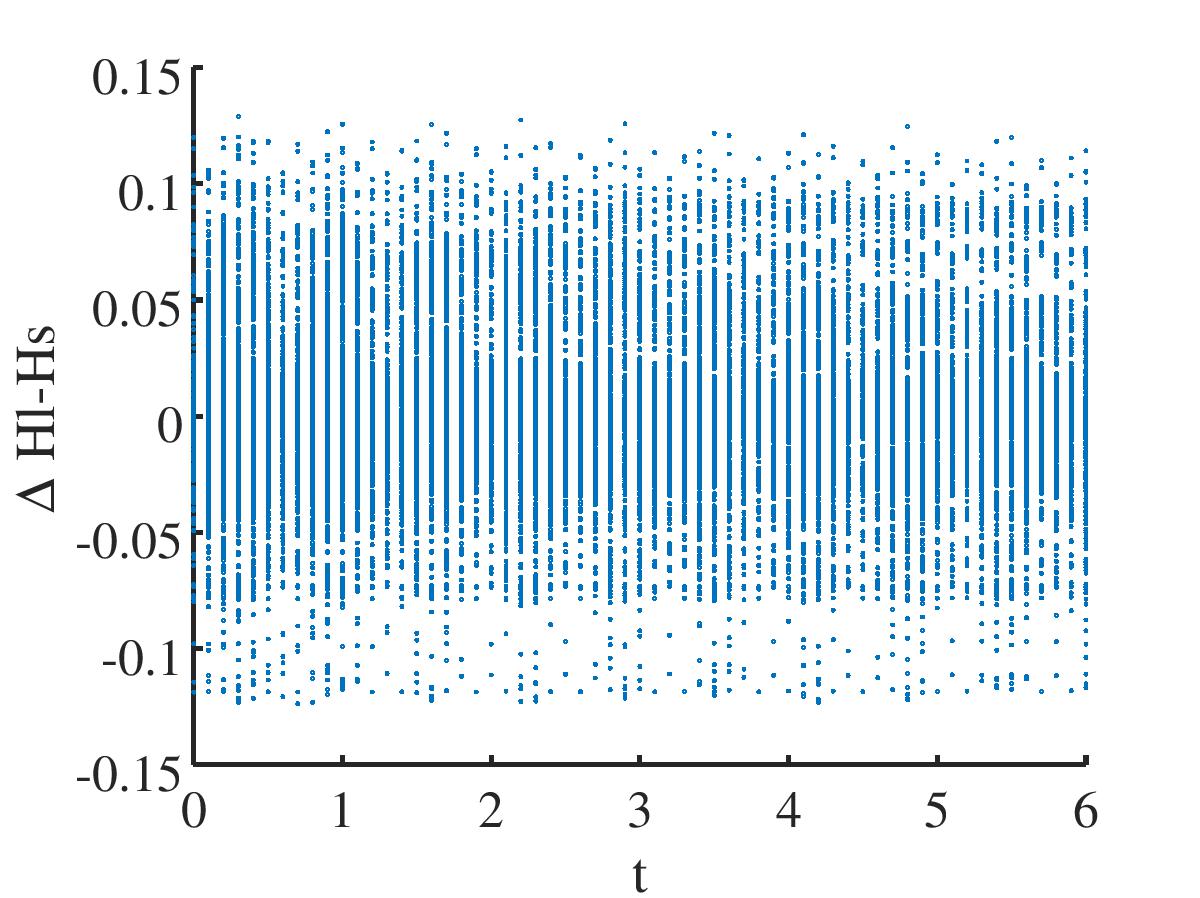}
\label{fig:side:a}

\end{minipage}%
\begin{minipage}[t]{0.5\linewidth}
\centering
\includegraphics[width=1.6in]{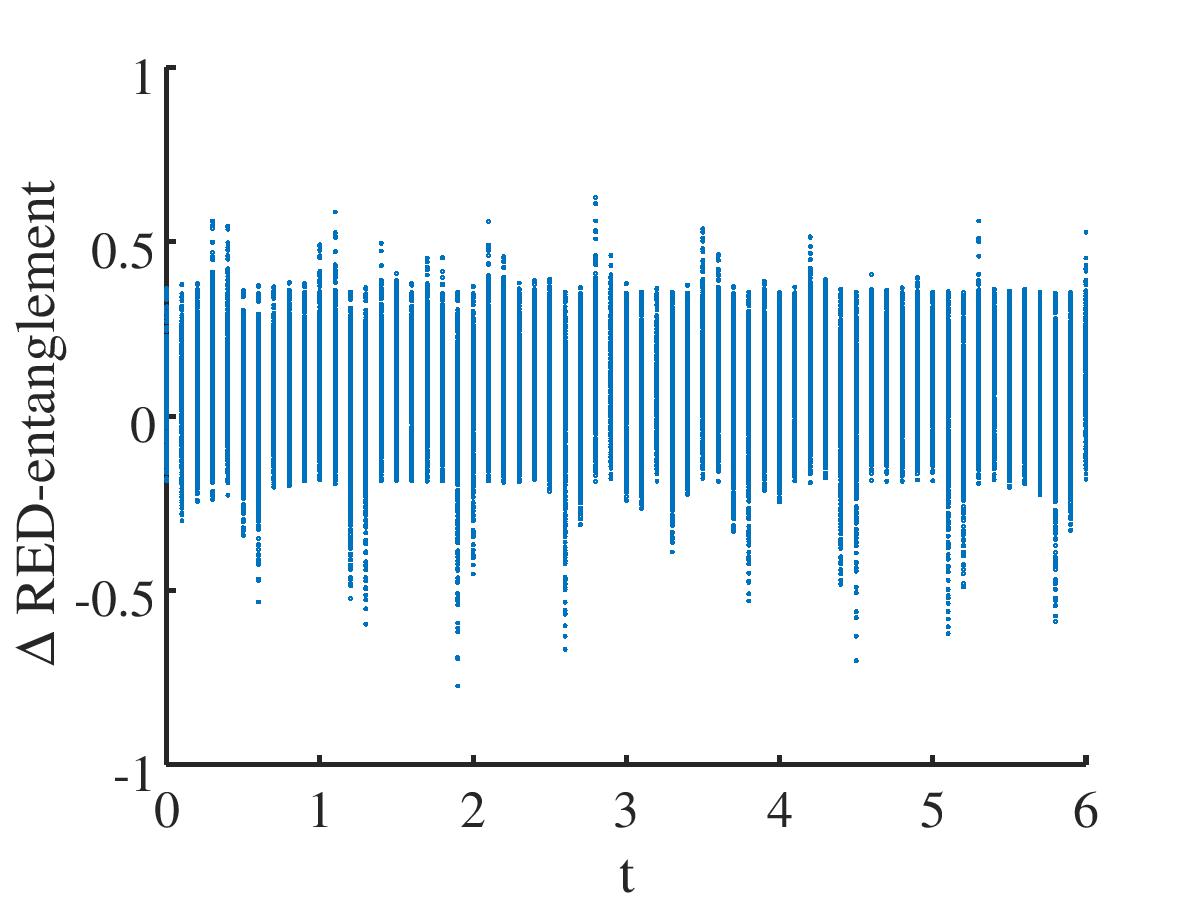}
\label{fig:side:b}
\end{minipage}

\setlength{\abovecaptionskip}{-0.2cm}
\caption{\label{fig:epsart} (Color online) The point diagram shows the difference between the different quantification methods at different time t for all the initial X states. The value of the point represents the value of the difference. (a) is difference between $D_{Br}$ and $D_{HL}$; (b) is difference between $D_{Br}$ and $D_{HS}$; (c) is difference between $D_{HL}$ and $D_{HS}$; (d) is difference between $RED$ and $concurrence$. }
\end{figure}

\emph{\textbf{The effect of the anisotropic coupling between qubits for freezing phenomenon.}}

As an interesting phenomenon in the process of quantum correlation evolution, the freezing phenomenon shows a robust feature of a family of two-qubit models
subject to nondissipative decoherence, which was first found for classical correlations [40]. Later, Mazzola, Piilo, and
Maniscalco [41] displayed the similar behavior for the quantum correlations, and Lang and Caves [42] provided a complete geometry picture explanation for Bell-diagonal states.

Later, some efforts have been devoted to discussing the condition for the frozen-discord [11,24-26] of other special states, including X states and SCI states [25-26,37]. These results demonstrate that the freezing conditions may be different for various forms of discord, which is also related to the study in the third part of this work. Here, as a supplement of our previous works [22], the effect of the anisotropic coupling parameters $J_{1}$ and $J_{2}$ for quantum correlation are shown in Fig.2. It reveals that the isotropic coupling is beneficial to the preservation of freezing phenomenon (see black line), with the amplitude of shock within the range $10^{-2}$ to $10^{-3}$. This property can be explained by the information transfer between the system and the environments. The anisotropic coupling system can arouse the different degrees of the information flow between system and the environments for different spin directions. Although there are some small differences (less than $10^{-2}$) in the evolution behavior of freezing phenomenon, the same state also shows freezing phenomenon under the same conditions, $\textit{e.g}$, the state ($a=\delta=0,b=0.4,c=0.5,\beta=0.4$). This property promotes the study about constructing the relationship between $D_{Br}$ and $RED$ which pave a way to further study the freezing condition for different quantification methods. Simultaneously, considering the results of our previous works [21-23,26,37], we choose the coupling parameter $q$ to generate the samples in the next item.

\begin{figure}
\begin{minipage}[t]{0.5\linewidth}
\centering
\includegraphics[width=1.7in]{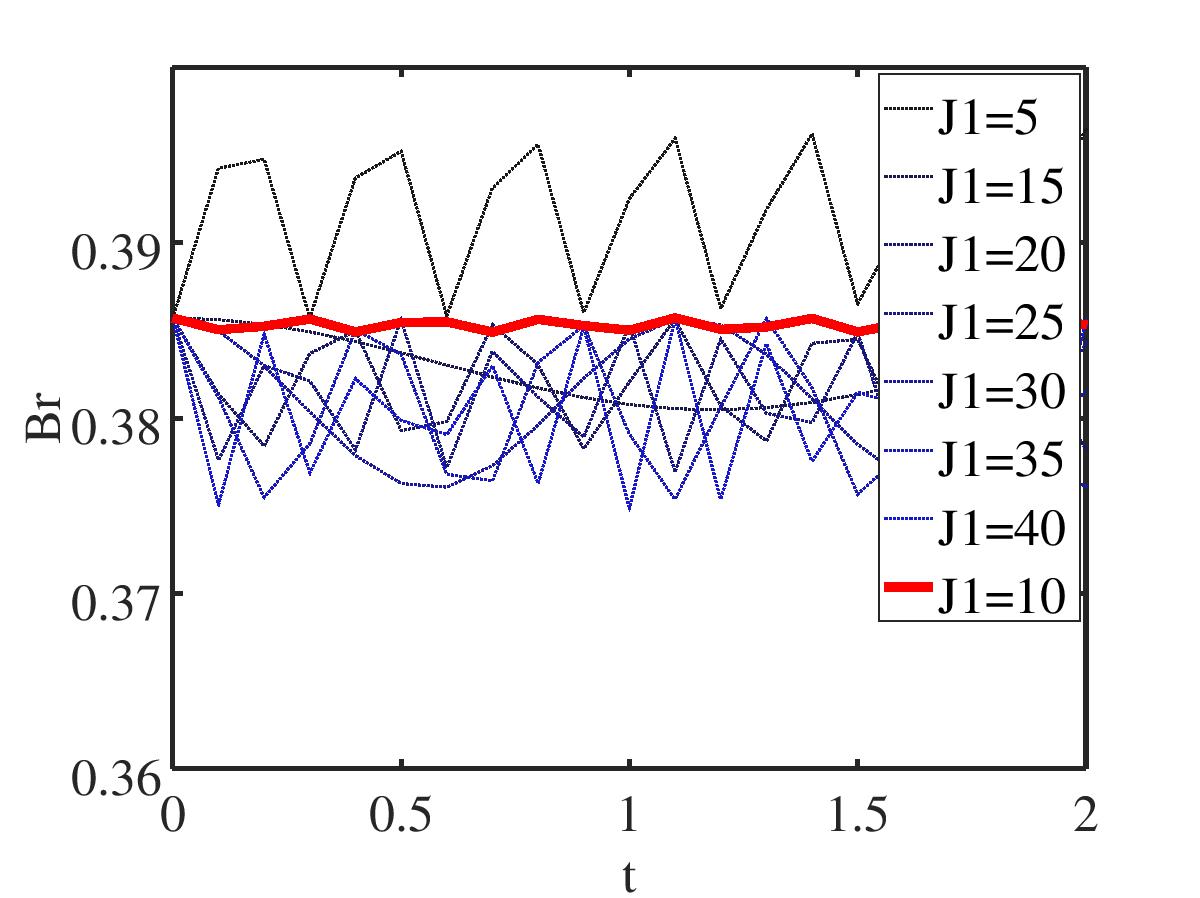}
\label{fig:side:a}
\end{minipage}%
\begin{minipage}[t]{0.5\linewidth}
\centering
\includegraphics[width=1.7in]{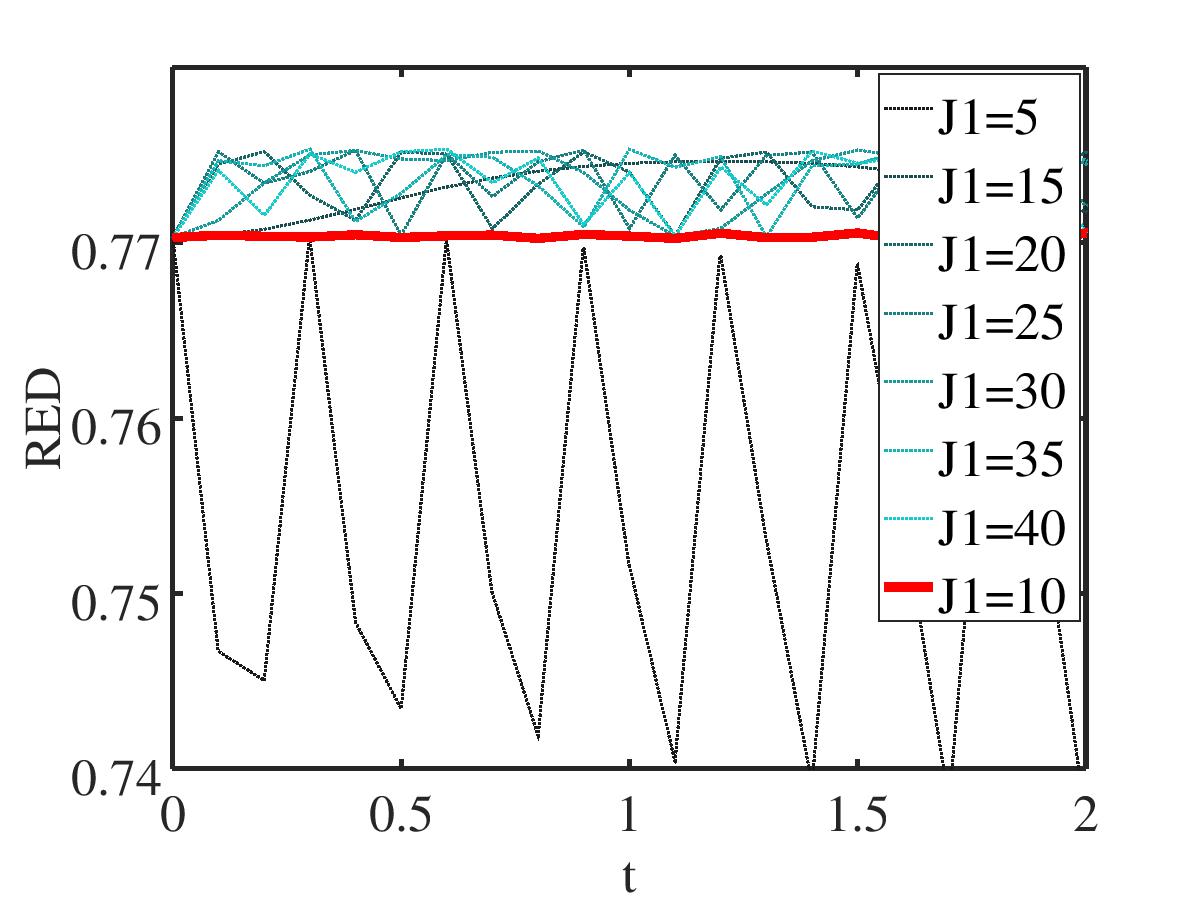}
\label{fig:side:b}
\end{minipage}
\setlength{\abovecaptionskip}{-0.2cm}
\caption{\label{fig:epsart} (Color online) The effect of the anisotropic coupling parameter for freezing phenomenon under different $J_{1}$. The quantum state corresponds to the parameters   $a=\delta=0,b=0.4,c=0.5,\beta=0.4$, $J_{2}=10ps^{-1}$. The other parameters are the same as Fig. 1.}
\end{figure}

\emph{\textbf{Machine learning study of the relationship between the $D_{Br}$ and RED($\alpha=2$).}}

From the point of view of invariance of physical laws, even for different methods, the same physical problem should have the same result. So there are some relations between the geometric and  entropy style discords.

Cianciaruso $\textit{et.al}$ discussed the geometric measure of dicord-type correlations based on the Bures distance ($d_{Bu}$)[11], which is defined as follows:
\begin{equation}
D_{Br}\equiv\inf_{\chi'}d^{2}_{Bu}(\rho,\chi')=\inf_{\chi'} 2(1-Tr([\sqrt{\chi'}\rho\sqrt{\chi'}]^{1/2}))
\end{equation}
where the set of classical-quantum states $\chi'=\sum_{i}p_{i}|i><i|^{A}\otimes \omega^{B}_{i}$, $p_{i}$ is a probability distribution, $\{|i>^{A}\}$ denotes an orthogonal basis for subsystem A, $\omega^{B}_{i}$ is an arbitrary ensemble of states for subsystem B, and $d_{Bu}(\rho,\chi')$ is the Bures distance.

\quad Because it is difficult to obtain mathematically analytic form of Eq.(6) for general models,  some numerical calculation methods were proposed in ref.[33] which are also adopted in this work to study $D_{Br}$ based on the relation between quantum Fisher information and the Bures distance. The Bures distance can be rewritten
\begin{eqnarray}
\mathcal{P}^{A}(\rho_{AB}|\Gamma)&=&\frac{1}{4}\mathop{min}_{H_{A}^{\Gamma}}F(\rho_{AB};H_{A}^{\Gamma})
\end{eqnarray}
where $F$ denotes the quantum Fisher information, $F(\rho_{AB};H_{A}^{\Gamma})=4\sum_{i<k:q_{i}+q_{k}\neq0}\frac{(q_{i}+q_{k})^{2}}{q_{i}+q_{k}}|\langle \psi_{i}|(H_{A}^{\Gamma}\otimes \mathbb{I}_{B})|\psi_{k}\rangle|^{2}$, with ${q_{i}, |\psi_{i}\rangle}$ denoting respectively the eigenvalues and eigenvectors of $\rho_{AB}$, and the minimum is taken over the set of all local Hamiltonians $H_{A}^{\Gamma}$.

The R\`{e}nyi quantum discord of $\rho_{AB}$ is an extension of quantum discord and is defined for $\alpha\in(0,1)\cup(1,2]$ as follows [29-30]
\begin{eqnarray}
D_{\alpha}(\rho_{AB})=\mathop{inf}_{\Pi_{k}^{A}}I_{\alpha}(E;B\mid X)_{\tau_{XEB}}
\end{eqnarray}
where the  R\`{e}nyi conditional mutual information $I_{\alpha}(E;B\mid X)_{\tau_{XEB}}$ satisfies:
\begin{eqnarray}
I_{\alpha}(E;B\mid X)_{\tau_{XBE}}&=&\frac{\alpha}{\alpha-1}\log Tr\{(\rho_{X}^{\frac{\alpha-1}{2}}Tr_{E}\{\rho_{EX}^{\frac{1-\alpha}{2}}\rho_{EBX}^{\alpha}\nonumber\\
&&\rho_{EX}^{\frac{1-\alpha}{2}}\}\rho_{X}^{\frac{\alpha-1}{2}})^{\frac{1}{\alpha}}\}
\end{eqnarray}

where the the classical output $X$ denotes the measurement acting on system $A$ and $E$ is an environment for the measurement map [30]. In this paper, we choose the von Neumann measurement ${\Pi_{i'}=|i'\rangle\langle i'| (i=0,1)}$ with  two angular  parameters $\theta$ and $\phi$: $|0'\rangle=\cos(\theta/2)|0\rangle+e^{i\phi}\sin(\theta/2)|1\rangle$ and $|1'\rangle=\sin(\theta/2)|0\rangle-e^{i\phi}\cos(\theta/2)|1\rangle$ ($0\leq\theta\leq\pi/2;0\leq\phi\leq\pi$). The properties of the R\`{e}nyi quantum discord are shown in Table 2 of Ref.[30].

Here, there are two points that arouse our interests about studying the relationship between $D_{Br}$ and $RED$ for $\alpha=2$. One is some advantages of $D_{Br}$ over other geometric discords, such as the ordering of quantum correlation [11,33], convex, monotonous [29-30] and the description of the freezing phenomenon [26]. The other is that there is similar structure $Tr(.)$ which relates to the definition of fidelity for $D_{Br}$ and $RED$ of $\alpha=2$. Unfortunately, it is very difficult to get analytical solution because of the nonlinear definitions of $D_{Br}$ and $RED$ and the structural dependence of the density matrix. So we apply a neural network to search the relationship between $D_{Br}$ and $RED$ for $\alpha=2$.

Since the multi-layer neural network can simulate the random function, we used the neural network model to construct the relationship between $D_{Br}$ and $RED$ based on the data of $D_{Br}$ and $RED$. Artificial neural networks is a multi-layer perception that inspired by the biological neural networks that constitute animal brains. It is a network of simple unit which also named neuron. Each neuron is defined as a non-linear mapping from sum of its inputs to output. Through training process, the weight parameters of units are adjusted so that the neural network extracts the potential patterns in data sets [43]. Fig.3 shows a structure graph of a general neural network. It has input data $x_{i}$, hidden layer neural $a_{i}^{l}$ and output data $y$, satisfying
\begin{figure}
\begin{minipage}[t]{1\linewidth}
\centering
\includegraphics[width=2.2in]{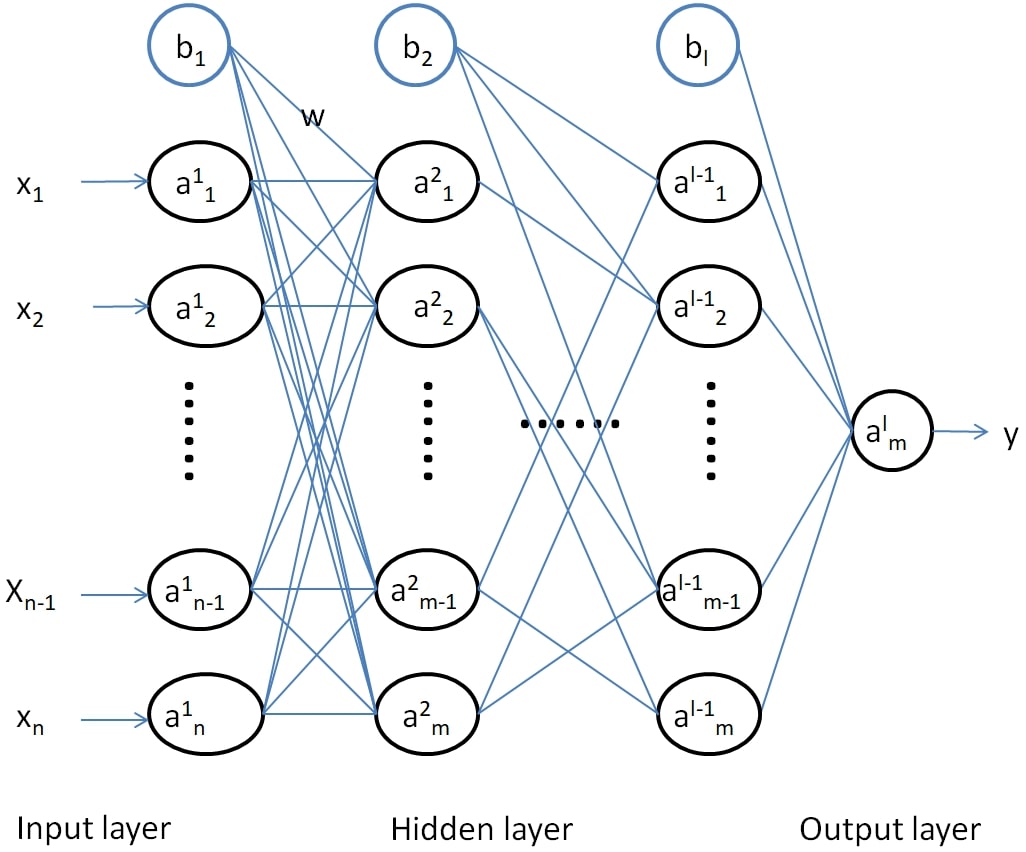}
\label{fig:side:a}
\end{minipage}%
\setlength{\abovecaptionskip}{-0cm}
\caption{\label{fig:epsart} (Color online) The structure graph of a general neural network. $x_{i}$ ($y$) is input(output) data, $b_{l}$ is the bias unit of $lth$ layer, $ a^{(l)}_{i} $ is the $ith$ unit of $l$ layer and the input layer satisfies $l=1$. $ W^{(l)} $ is the weight matrix, the elements $ w^{(l)}_{ij} $ is the weight of connection between $ a^{(l-1)}_{j} $ and  $ a^{(l)}_{i} $.
}
\end{figure}
\begin{eqnarray}
\textbf{z}^{l}&=&W^{(l)}\textbf{A}^{l-1}+\textbf{b}_{l-1}\nonumber\\
a_{i}^{l}&=&f(z_{i}^{l})
\end{eqnarray}
where the $l$th layer neural cells denote $\textbf{A}^{l}=[a_{1}^{l},a_{2}^{l}....a_{n}^{l}]$ and for $l=1$, $\textbf{A}^{1}=[x_{1}^{l},x_{2}^{l}....x_{n}^{l}]$, $\textbf{z}^{l}=[z_{1}^{l},z_{2}^{l}....z_{n}^{l}]$.

In order to realize the nonlinear relation between input and output of each neural node, the activation function $F(z)$ is required
\begin{equation}
f(z)=\emph{tanh}(z)=\frac{e^{z}-e^{-z}}{e^{z}+e^{-z}}
\end{equation}

We adjust the parameters of neural network to minimize the cost function by using back-propagation algorithm and gradient descent method.
\begin{equation}
cost(x)=\sum(y-y')^{2}
\end{equation}
here, summation for all the training data(training samples). $y'$ and $y$ denote the predicted value and real value of $D_{Br}$.

We constructed $4$ layer neural network for our problem. The number of neurons per layer is $7,13,1,1$. The learning process of neural network is show in  Algorithm $1$.

\begin{algorithm}[H]
\caption{Learning process of neural network}
\label{alg1}
\begin{algorithmic}[1]
\STATE Input: matrix $n*7$ \($n$ data samples with $7$ features\)
\STATE Output: matrix $n*1$, the predicted value $y'$ of $D_{Br}$ for each sample)
\STATE $1.$  Initial the parameters ($\textbf{W,b}$) in neural network.
\STATE $2.$ Split the data set into training data, validation data and test data randomly with proportion $60\%, 20\%$ and $20\%$.
\STATE for t=1:100000 do
\STATE Minimize the difference between predicted value $y'$ and real value $y$ by updating the parameters using gradient descent on training data
\STATE End for
\STATE $3.$ use the validation data to choose the neural network with the minimal cost function
\end{algorithmic}
\end{algorithm}

$\emph{Samples}--$ Because the matrix form of $\rho_{s}(t)$ is also X form for $X$ initial state, the $X$ states samples are generated from the data of $\rho_{s}(t)$ for $60$ samples per $6$ seconds. Similarly, we change the parameter $q$ to generate another group of samples in the same process. The total number of samples are more than one hundred and twenty thousand with the repetition rate less than $1\%$. Based on the Eq.(7) and (9), the values of $D_{Br}$ and $RED(\alpha=2)$ are obtained for these samples.

$\emph{Feature}--$ Based on Eq.(6)-(7) and (9), We choose seven parameters, including the four eigenvalues of $\rho_{s}(t)$ and $\theta$ and $\phi$ which are introduced in $RED$ calculation process, and $RED$ as the input features of neural network. Data analysis reveals an important character of data which is classified to $\theta =0$ and $\theta =\frac{\pi}{4}$.

$\emph{Neural network model}--$ A bridge (relationship) has been built between $RED(\alpha=2)$ and $D_{Br}$ for above two classifications. Here, we randomly choose $60\%$ of data as training data, $20\%$ as validation data and $20\%$ as test data. In Fig.(4), the red line shows that at the end of training, the mean-square error (MSE)which is equal to the expectations of $cost$ rapidly decreases at first hundreds epochs and eventually converges after hundred thousand epochs. For $\theta = 0$($\theta = \pi/4$), the MSE is less than $0.004$($0.0027$). This means that a good relationship is constructed based on our model.

\begin{figure}
\begin{minipage}[t]{0.5\linewidth}
\centering
\includegraphics[width=1.8in]{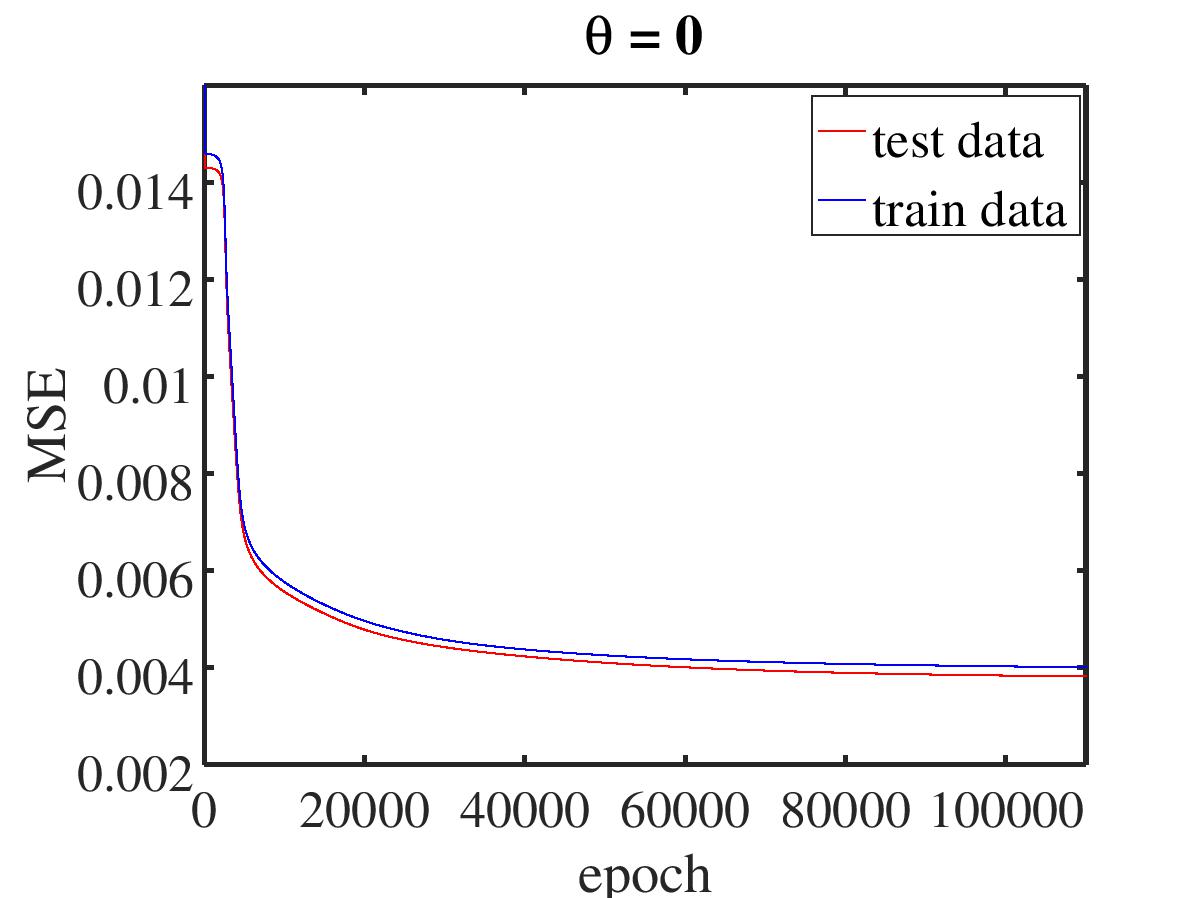}
\label{fig:side:a}
\end{minipage}%
\begin{minipage}[t]{0.5\linewidth}
\centering
\includegraphics[width=1.8in]{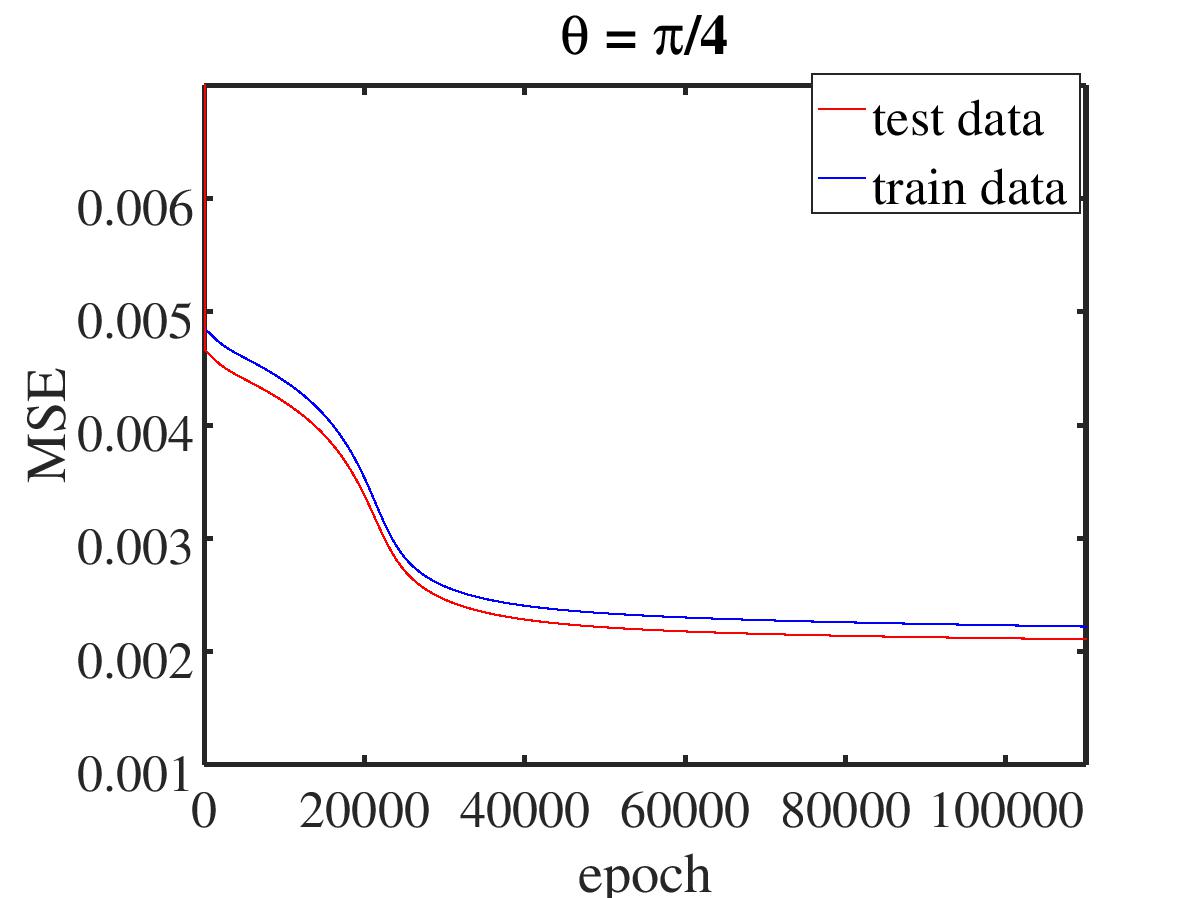}
\label{fig:side:b}
\end{minipage}
\setlength{\abovecaptionskip}{-0.2cm}
\caption{\label{fig:epsart} (Color online) The figure of the mean-square error (MSE) which is equal to the expectations of $cost$ change with epoch for $\theta = 0$ and $\theta = \pi/4$, respectively. The red(blue) line shows the behavior of test(train) data. The other parameters are the same as Fig. 1.}
\end{figure}

\emph{Overfitting--} For all machine learning applications, the training process should be carefully designed to avoid overfitting. Dropout is a regularization technique to reduce  overfitting in neural networks by preventing complex co-adaptations on training data. This method is applied to prevent overfitting, which means that we temporarily remove some units from the network, along with all its incoming and outgoing connections. Meanwhile, we randomly choose units to drop at each epoch [44].

 In Fig.(4), the two lines are rapidly decreasing in the first few hundred iterations, and then gradually converge to zero. Meanwhile, comparing the blue and red lines, it is shown that the parameters that applied to the training set data are also applicable to the test set as the distance between two lines is very small (in the magnitudes $10^{-4}$). That is to say, the parameters of the neural network can be generalized without overfitting. It also further demonstrates that the general relationship between $RED(\alpha=2)$ and $D_{Br}$ for X states is correct. Remarkably, our results not only hold for other systems, but also pave a way for the further study of the physical nature of quantum correlation.

 Finally, from the physical perspective, the quantum correlation shows the different characteristics of the quantum states contrasting with the classical
states or the changing degree of the quantum states when it suffers the local disturb. So, the system information presented by different discordlike definitions
will be different. Searching the link between these defines, it will not only help us to understand the differences and commonalities of systematic information obtained by different definitions, but also help us to understand the total properties of quantum states, such as coherence, and the properties of entanglement.

\emph{\textbf{Conclusion$-$}}
In this paper, two main results about quantum correlation are presented. One is that the ordering of the value of quantum correlation is obtained with different quantization methods for an open quantum system scenario. It also hints that $D_{Br}$ and $RED$ quantization methods maybe better than the others methods. In addition, the anisotropic coupling between qubits can affect the freezing phenomenon. The other is that machine learning method is firstly applied to study quantum correlation and successfully construct the relationship between geometric ($D_{Br}$) and entropy ($RED$) style discord for $X$ form states. This bridge will help to study the difference of quantum correlation between different quantization methods.

\end{document}